\documentclass{article}
\usepackage{amssymb}

\begin{document}

\title{Discrete Randomness in Discrete Time Quantum Walk: Study via
Stochastic Averaging }
\author{D. Ellinas$^{\ast }$, A. J. Bracken$^{\#}$ and I. Smyrnakis$^{\$}$ \\
$^{\ast }$Technical University of Crete Department of Sciences \\
$M\Phi Q$ Research Unit GR-731 00 Chania Crete Greece\\
\texttt{ellinas@science.tuc.gr}\\
$^{\#}$Centre for Mathematical Physics, Department of\\
Mathematics, University of Queensland, Brisbane 4072 Australia \\
\texttt{ajb@maths.uq.edu.au}\\
$^{\$}$Technological Education Institute of Crete, P.O. Box 1939,\\
GR-71004, Heraklion, Greece\\
\texttt{smyrnaki@tem.uoc.gr}}
\maketitle

\begin{abstract}
The role of classical noise in quantum walks (QW) on integers is
investigated in the form of discrete dichotomic random variable affecting
its reshuffling matrix parametrized as a $SU2)/U(1)$ coset element. Analysis
in terms of quantum statistical moments and generating functions, derived by
the completely positive trace preserving (CPTP) map governing evolution,
reveals a pronounced eventual transition in walk's diffusion mode, from a
quantum ballistic regime with rate $\mathcal{O}(t)$ to a classical diffusive
regime with rate $\mathcal{O}(\sqrt{t}),$ when condition $($\textit{strength
of noise parameter}$)^{2}\times ($\textit{number of steps}$)=1,$ is
satisfied. The role of classical randomness is studied showing that the
randomized QW, when treated on the stochastic average level by means of an
appropriate CPTP averaging map, turns out to be equivalent to a novel
quantized classical walk without randomness. This result emphasizes the dual
role of quantization/randomization in the context of classical random walk.
Keywords: Quantum walk, randomness, CP map, quantization
\end{abstract}

\section{\textbf{\protect\bigskip }\protect\bigskip Introduction}

\textit{Quantum Walk on }$%
\mathbb{Z}
:$ \textit{Recapitulation.} The essential feature of a QW on a line is the
promotion of the mathematical correspondence: \textit{left/right}$%
\rightarrow $\textit{\ head/tails, }between \textit{walker}'s move
directions and \textit{coin}'s sides, to a dynamic interaction among two
physical quantum systems\cite{yaharonov}-\cite{watrous}. This is realized by
introducing state Hilbert spaces $H_{\mathrm{w}}=$span$(|m\rangle )_{m\in
Z}\thickapprox
\mathbb{C}
^{%
\mathbb{Z}
}$ and $H_{\mathrm{c}}=$span$(|+\rangle ,|-\rangle )\thickapprox
\mathbb{C}
^{2},$ for the quantum walker and coin systems respectively. In $H_{\mathrm{w%
}} $ the algebra Euclidean group $ISO(2)$ with generators $\{\widehat{L},%
\widehat{E}_{+},\widehat{E}_{-}\}$ satisfying relations $[\widehat{L},%
\widehat{E}_{\pm }]=\pm \widehat{E}_{\pm },$ $[\widehat{E}_{+},\widehat{E}%
_{-}]=0,$is represented by the \textit{step operators }$\widehat{E}_{\pm
}|m\rangle =|m\pm 1\rangle $ and the \textit{position operator }$\widehat{L}%
|m\rangle =m|m\rangle .$ \ Introduce also the coin space projection
operators $P_{\pm }=|\pm \rangle \langle \pm |$. One step of a CRW is
described by the unitary $V_{\mathrm{cl}}=P_{+}\otimes \widehat{E}%
_{+}+P_{-}\otimes \widehat{E}_{-}$, acting on the total space $H_{\mathrm{c}%
}\otimes H_{\mathrm{w}}$, and via its conditional action on walker states,
realizes the walker's move. Explicitly, coin $\rho _{\mathrm{c}}$\ and
walker $\rho _{\mathrm{w}}$ density matrices, initially in a product state $%
\rho _{\mathrm{c}}\otimes \rho _{\mathrm{w}}$, interact unitarily via the
map $\rho _{\mathrm{c}}\otimes \rho _{\mathrm{w}}\rightarrow V_{\mathrm{cl}%
}\rho _{\mathrm{c}}\otimes \rho _{\mathrm{w}}V_{\mathrm{cl}}^{\dagger }=$Ad$%
V_{\mathrm{cl}}(\rho _{\mathrm{c}}\otimes \rho _{\mathrm{w}})$ (written
alternatively in terms of the adjoint action Ad$X(\rho ):=X(\rho )X^{\dagger
})$, and then de-couple by an unconditional quantum measurement of the coin
subsystem, realized by the partial trace Tr$_{\mathrm{c}}$, \emph{i.e.} $V_{%
\mathrm{cl}}\rho _{\mathrm{c}}\otimes \rho _{\mathrm{w}}V_{\mathrm{cl}%
}^{\dagger }\rightarrow $Tr$_{\mathrm{c}}(V_{\mathrm{cl}}\rho _{\mathrm{c}%
}\otimes \rho _{\mathrm{w}}V_{\mathrm{cl}}^{\dagger })$. This constitutes a
dynamic realization of the coin tossing process. The diagonal elements of
the resulting walker's density matrix $\mathcal{E}_{\mathrm{cl}}(\rho _{%
\mathrm{w}})=$Tr$_{\mathrm{c}}($Ad$V_{\mathrm{cl}}(\rho _{\mathrm{c}}\otimes
\rho _{\mathrm{w}})),$\ realizes the CRW on the integers \cite{bet},\cite%
{ellsmyrnEpsilon}.

Quantization of CRW is conceived as the incorporation in $H_{\mathrm{c}}$ of
an additional unitary operator $U$, the coin reshuffling matrix, in two
different modes: either via the map Ad$V_{\mathrm{cl}}\rightarrow $Ad$V_{%
\mathrm{cl}}\circ $Ad$U\otimes \mathbf{1,}$ resulting into the $U$-\textit{%
quantization rule,} or via the map Ad$V_{\mathrm{cl}}\rightarrow $Ad$V_{%
\mathrm{cl}}\circ \mathcal{E}_{\mathrm{c}}\mathbf{:=}$Ad$V_{\mathrm{cl}%
}\circ \sum_{i}\lambda _{i}$Ad$U_{i}\otimes \mathbf{1,}$ for given ensemble $%
\{\lambda _{i},U_{i}\}_{_{i}}$ of reshuffling matrices $U_{\mathrm{i}},$
forming a CPTP\cite{nielsenchuang}, quantization map $\mathcal{E}_{\mathrm{c}%
}\mathbf{:=}\sum_{i}\lambda _{i}$Ad$U_{i},$ resulting into the $E$-\textit{%
quantization rule }\cite{ellsmyrnEpsilon}.

\section{Noisy diagonal QW}

Consider the unitary `walker-coin' operator $V(s)=V_{\mathrm{cl}}U(s)\otimes
\mathbf{1}_{\mathrm{w}}\mathbf{,}$ or explicitly
\begin{equation}
V(s)=P_{+}\,U(s)\otimes \widehat{E}_{+}+P_{-}\,U(s)\otimes \widehat{E}_{-},
\end{equation}%
where the reshuffling matrix $U(s),$ according to the Appendix, is \ taken
to be an element of the coset $SU(2)/U(1)\thickapprox CP.$ Referring to (\ref%
{coset}), $\ $the coordinate of the complex projective variable $\theta \in
CP,\ $\ is chosen to be a dichotomous random variable (rv) $\theta
=s\epsilon ,$ $\epsilon >0,$ with values $\theta =\pm \epsilon ,$ $\ $where $%
s=\pm 1$, with equal probabilities. The reshuffling matrix
\begin{equation}
U(s)=\mathcal{N}\left(
\begin{array}{cc}
1 & s\epsilon \\
-s\epsilon & 1%
\end{array}%
\right) ,
\end{equation}%
with $\mathcal{N}\equiv \frac{1}{\sqrt{1+\epsilon ^{2}}},$ provides the
unitaries $V(s=\pm 1)$ that apply at random, with probability $q_{\pm }=1/2$
(unbiased walk), so that the `walker-coin' density operator evolves from the
$N$-th to $(N+1)$-th step as

\begin{equation}
\rho _{N+1}=\frac{1}{2}\sum_{s=\pm 1}V(s)\rho _{N}V(s)^{\dagger }.
\label{evolution}
\end{equation}

Rewrite this as $\rho _{N+1}=\widehat{\mathcal{E}}\mathcal{(}\rho _{N}%
\mathcal{)\equiv }\left\langle V(s)\rho _{N}V(s)^{\dagger }\right\rangle $ $=%
\frac{1}{2}\sum_{s=\pm }V_{q}(s)\rho _{N}V_{q}(s)^{\dagger }=\frac{1}{2}$Ad$%
V(s)\rho _{N},$ where $\widehat{\mathcal{E}}:$Lin$(H_{\mathrm{c}}\otimes H_{%
\mathrm{w}})\rightarrow $Lin$(H_{\mathrm{c}}\otimes H_{\mathrm{w}})$ stands
for a CPTP map acting on the composite coin-walker system which implements
the ensemble average of the random unitary evolution. This density matrix
iteration provides the sequence $\{\rho _{N}\}_{N=1}^{\infty },$ which
subsequently provides, for any observable operator $A_{\mathrm{c}}\otimes B_{%
\mathrm{w}}$ on $H_{\mathrm{c}}\otimes H_{\mathrm{w}},$ a sequence of
quantum moments \ $\{\mu _{N}\}_{N=1}^{\infty },$ as $\mu _{N+1}=$Tr$(\rho
_{N+1}A_{\mathrm{c}}\otimes B_{\mathrm{w}})\equiv \left\langle A_{\mathrm{c}%
}\otimes B_{\mathrm{w}}\right\rangle _{N}.$ Utilizing the state-observable
(trace inner product) duality, we derive the moments
\begin{equation}
\mu _{N+1}=\mathrm{{Tr}(\widehat{\mathcal{E}}\mathcal{(}\rho _{N})A_{{c}%
}\otimes B_{{w}})={Tr}(\rho _{N}\widehat{\mathcal{E}}^{\ast }\mathcal{(}A_{{c%
}}\otimes B_{{w}}))={Tr}(\rho _{1}\widehat{\mathcal{E}}^{N\ast }\mathcal{(}%
A_{{c}}\otimes B_{{w}})),}
\end{equation}%
either evolving the state $\widehat{\mathcal{E}}$, or the observables by the
dual CP map $\widehat{\mathcal{E}}^{\ast }.$ The latter has as Kraus
generators, the hermitean conjugates of those of $\widehat{\mathcal{E}}.$ By
means of the dual map $\widehat{\mathcal{E}}^{\ast }$, moment iterations for
various \ choices of observables $A_{\mathrm{c}}\otimes B_{\mathrm{w}},$ are
investigated next. \

Let $\rho _{N}=\sum_{\alpha \mathrm{,}\beta \in \{+,\,-\}}\sum_{k\mathrm{{,}%
l\in Z}}(\rho _{N})_{k\mathrm{{,}\alpha {,}l{,}\beta }}\left\vert \alpha
\right\rangle \left\langle \beta \right\vert \otimes \left\vert
k\right\rangle \left\langle l\right\vert ,$ be the density matrix describing
the state of the \textquotedblleft coin-walker" system, where $(\rho _{N})_{k%
\mathrm{{,}\alpha {,}l{,}\beta }}$ are elements labelling walker positions $%
k,l$ $\in
\mathbb{Z}
,$ and coin states $\alpha $, $\beta $ $\in \{+,\,-\}$. \ Assume initially a
factorized state $\rho _{0}=\rho _{0}^{(\mathrm{{c})}}\otimes \rho _{0}^{(%
\mathrm{{w})}}$, $\rho _{0\,}^{(\mathrm{{w})}}=\left\vert 0\right\rangle
\left\langle 0\right\vert ,$ $\rho _{0}^{(\mathrm{{c})}}=$diag$(\cos
^{2}\gamma ,\sin ^{2}\gamma ),$ for $\gamma $ $\in \lbrack 0,\pi )$. The
resulting matrix valued recurrence relation implies that $\rho _{N}$ is also
diagonal. Setting $(\rho _{N})_{\mathrm{{\thinspace }kk}}=$diag$(\alpha
_{Nk},\beta _{Nk}), $ for diagonal components obtained by expectation values
of the projectors,
\begin{eqnarray}
\alpha _{Nk} &=&\mathrm{{Tr}(\rho _{N}P_{+}\otimes P_{k})\equiv \left\langle
P_{+}\otimes P_{k}\right\rangle _{N},} \\
\beta _{Nk} &=&\mathrm{{Tr}(\rho _{N}P_{-}\otimes P_{k})\equiv \left\langle
P_{+}\otimes P_{k}\right\rangle _{N}.}
\end{eqnarray}%
leads to recurrence relations%
\begin{eqnarray}
\alpha _{N+1\mathrm{{\thinspace }k}} &=&\mathcal{N}^{2}\left( \alpha
_{N\,k-1}+\epsilon ^{2}\beta _{N\,k-1}\right) \,, \\
\beta _{N+1\,k} &=&\mathcal{N}^{2}\left( \epsilon ^{2}\alpha _{N\,k+1}+\beta
_{N\,k+1}\right) \,,  \label{recurrence1}
\end{eqnarray}%
for $N=0,\,1,\,2,\,\dots $, $k\in
\mathbb{Z}
$, with initial conditions $\alpha _{00}=\cos ^{2}\gamma $, $\beta
_{00}=\sin ^{2}\gamma $, $\alpha _{0\,k}=\beta _{0\,k}=0$ for $k\neq 0$.

\section{Recurrence relations for moments}

The recurrence relations (\ref{recurrence1}) are difficult to solve exactly,
except when $\epsilon =0$ or $1$. We consider instead the \emph{quantum}
\emph{moments} of the distribution of probability over $k$-values (`walker
positions') after $N$ iterations. Accordingly, let $A_{N}^{(p)}=\sum_{k=-%
\infty }^{\infty }\alpha _{N\,k}\,k^{p}\,,$ and $B_{N}^{(p)}=\sum_{k=-\infty
}^{\infty }\beta _{N\,k}\,k^{p}\,,$for $p=0,\,1,\,2,\,\dots $. Then $%
A_{N}^{(p)}$ is the $p$-th moment of the distribution of probability over $k$
values, with coin spin `up,' and $B_{N}^{(p)}$ is the $p$-th moment of the
distribution of probability over $k$ values, with coin spin `down.' The sum $%
A_{N}^{(p)}+B_{N}^{(p)}$ is the total $p$-th moment of the distribution of
probability over $k$ values. The operator expression of these moments and
their combinations $S_{N}^{(p)}=A_{N}^{(p)}+B_{N}^{(p)}\,,$ and$\quad
D_{N}^{(p)}=A_{N}^{(p)}-B_{N}^{(p)},$ read%
\begin{eqnarray}
A_{N}^{(p)} &=&\mathrm{{Tr}(P_{+}\otimes \widehat{L}^{p}\rho
_{N}),B_{N}^{(p)}={Tr}(P_{-}\otimes \widehat{L}^{p}\rho _{N}),} \\
D_{N}^{(p)} &=&\mathrm{{Tr}((P_{+}-P_{-})\otimes \widehat{L}^{p}\rho _{N}),}
\\
S_{N}^{(p)} &=&\mathrm{{Tr}((P_{+}+P_{-})\otimes \widehat{L}^{p}\rho _{N})={%
Tr}(\mathbf{1}_{{c}}\otimes \widehat{L}^{p}\rho _{N})=:{Tr}(\widehat{L}%
^{p}\rho _{N}^{({w})}).}
\end{eqnarray}

The resulting recurrence relations for the moments are also difficult to
solve in closed form for general $N$ and $p$, except small values of $p$. In
this way we find that
\begin{equation}
A_{N}^{(0)}=\frac{1}{2}\left( 1+r^{N}cos(2\gamma )\right) \,,\quad
B_{N}^{(0)}=\frac{1}{2}\left( 1-r^{N}cos(2\gamma )\right) \,.
\label{zeroth_moments}
\end{equation}

For the higher moments, we find%
\begin{equation}
D_{N}^{(1)}=(1-r^{N})/(1-r)\,S_{N}^{(1)}=r cos(2\gamma
)[1-r^{N}]/[1-r]\,,  \label{first_moments}
\end{equation}%
and \
\begin{eqnarray}
D_{N}^{(2)} &=&2r cos(2\gamma )(1-r^{N})/(1-r)^{2}-Nr^{N} cos%
(2\gamma )(1+r)/(1-r) \\
S_{N}^{(2)} &=&N(1+r)/(1-r)-2r(1-r^{N})/(1-r)^{2}.
\end{eqnarray}

The second\ total moment \emph{i.e.} the expectation value of the square of
the position operator $\widehat{L},$ can also be written as%
\begin{eqnarray}
S_{N}^{(2)} &=&\mathrm{{Tr}(\widehat{L}^{2}\rho _{N}^{(w)})\equiv
\left\langle \widehat{L}^{2}\right\rangle _{N}} \\
&=&\frac{1}{2\epsilon ^{4}}(2N\epsilon ^{2}-1+\epsilon ^{4}-(1-\epsilon
^{2})^{N+1}/(1+\epsilon ^{2})^{N-1}).
\end{eqnarray}

There are two regimes of interest for the second moment, corresponding to
two value ranges of discrete time steps $N$ relative to the strength of
randomness coefficient $1/\epsilon ^{2}$. The first regime is when $%
1/\epsilon ^{2}\gg N\gg 1,$ and then
\begin{equation}
(1-\epsilon ^{2})^{N+1}/(1+\epsilon ^{2})^{N-1}=1-2N\epsilon
^{2}+(2N^{2}-1)\epsilon ^{4}+\mathcal{O}((N\epsilon ^{2})^{3}),
\label{expansion1}
\end{equation}%
so that
\begin{equation}
S_{N}^{(2)}=N^{2}+\frac{1}{2\epsilon ^{4}}\mathcal{O}((N\epsilon
^{2})^{3})\,.  \label{regime2}
\end{equation}

This is the `ballistic' or `inertial' regime, where the growth rate of the
second moment with $N$ is $\sqrt{\left\langle \widehat{L}^{2}\right\rangle
_{N}}\sim \mathcal{O(}N),$ that is characteristic of a quantum walk (QW)
\cite{ambainis}.

The second regime is when $N\gg 1/\epsilon ^{2}\gg 1$, and then%
\begin{equation}
S_{N}^{(2)}=\frac{N}{\epsilon ^{2}}\,\left\{ 1+\mathcal{O}(1/(N\epsilon
^{2})\right\} ,  \label{regime1}
\end{equation}%
\emph{i.e.} $\sqrt{\left\langle \widehat{L}^{2}\right\rangle _{N}}\sim
\mathcal{O}(\sqrt{N}),$ showing the typical growth rate of a CRW or
diffusion process. Note that, however small the $\epsilon ^{2}$, eventually $%
N$ becomes so large that the second regime is reached. The changeover occurs
for $N\epsilon ^{2}\approx 1$. Thus we see that for very small $\epsilon
^{2} $, the process behaves --- as far as the second moment is concerned ---
initially like a QW, but eventually like a CRW. This is precisely the
behavior observed in numerical simulations of a variety of quantum walks
with unitary noise \cite{biham} and in previous studies \emph{e.g.} \cite%
{buz}-\cite{silbernhorn}, of which the present process may be considered a
simple, special case, with the advantage that it is amenable to a complete
mathematical analysis.

\section{Randomization and Quantization of Walks}

Let us return to the evolution equation (\ref{evolution}) \ \ for the total
system density matrix. The following elaboration is possible; factoring the
evolution operator as $V(s)\equiv V_{\mathrm{cl}}U(s)\otimes \mathbf{1}_{%
\mathrm{w}}$ makes explicit that the stochastic averaging amounts to
averaging with respect to the matrices $U(s=+),$ $U(s=-)$ with probabilities
$p_{\pm }=\frac{1}{2},$ which is implemented by a CPTP map $\mathcal{E}_{%
\mathrm{c}}=\frac{1}{2}\sum_{s=\pm }$Ad$(U(s)\otimes \mathbf{1}_{\mathrm{w}%
}),$ acting non trivially only on $H_{\mathrm{c}}$. Explicitly \ we obtain $%
\rho _{\mathrm{{N}+1}}=\widehat{\mathcal{E}}\mathcal{(}\rho _{\mathrm{N}}%
\mathcal{)=}$ $V_{\mathrm{cl}}\mathcal{E}_{\mathrm{c}}\mathcal{(}\rho _{%
\mathrm{N}})V_{\mathrm{cl}}^{\dagger }=$Ad$V_{\mathrm{cl}}\circ \mathcal{E}_{%
\mathrm{c}}\mathcal{(}\rho _{\mathrm{N}}).$

Noticeable, the averaging map $\widehat{\mathcal{E}}=$Ad$V_{\mathrm{cl}%
}\circ \mathcal{E}_{\mathrm{c}}$ can be identified to be itself the CPTP
evolution map of another QW. This new QW will have as its \textquotedblleft
walker space" the total Hilbert space of the initial walk \emph{i.e.} $H_{%
\mathrm{t}}:=H_{\mathrm{c}}\otimes H_{\mathrm{w}}.$ In addition to the
operator sum decomposition of the map $\widehat{\mathcal{E}}$ \ defined
following \ref{evolution}, it could also be described by a unitary dilation
matrix $Y$ in a extended space $H_{\mathrm{aux}}\otimes H_{\mathrm{c}%
}\otimes H_{\mathrm{w}},$ where now an auxiliary vector space $H_{\mathrm{aux%
}}$ should be attached to the initial space $H_{\mathrm{t}},$ as an
additional "coin space". \ Such particular QW-like unitary dilation for $%
\widehat{\mathcal{E}}$ \ is given in terms of the unitary operator $%
Y=\sum_{s=\pm }P_{s}Q\otimes V(s),$ $\ $or $Y=Y_{\mathrm{cl}}Q\otimes
\mathbf{1}_{\mathrm{t}}\mathbf{,}$ acting on $H_{\mathrm{aux}}\otimes H_{%
\mathrm{c}}\otimes H_{\mathrm{w}}.$ Under fairly general conditions, for CP
maps with unitary Kraus generators (and $\widehat{\mathcal{E}}$ provides an
example), such a QW-like unitary dilation it is always possible (cf. \cite%
{Ellinas}, for a proof and related discussion). \ Explicitly for $\widehat{%
\mathcal{E}}$ we have that%
\begin{equation}
\rho _{N+1}=\widehat{\mathcal{E}}(\rho _{N})=\mathrm{{Tr}_{{aux}}{Ad}Y(\rho
_{{aux}}\otimes \rho _{{N}})={Tr}_{{aux}}{Ad}Y_{cl}\circ ({Ad}Q\otimes
\mathbf{1}_{{t}}\mathbf{)}(\rho _{{aux}}\otimes \rho _{N}).}
\end{equation}

Let $\rho _{\mathrm{aux}}=\left\vert a\right\rangle \left\langle
a\right\vert $ be auxiliary system's state with $\left\vert a\right\rangle $
a basis vector. The non-unique unitary $Q$ is chosen so that $q_{\pm
}=\left\langle \pm \right\vert Q\circ Q^{\ast }\left\vert a\right\rangle $,
\emph{i.e.} \newline
$Q=\left(
\begin{array}{cc}
\sqrt{q_{+}} & \sqrt{q_{-}} \\
-\sqrt{q_{-}} & \sqrt{q_{+}}%
\end{array}%
\right) .$ If $q_{\pm }=\frac{1}{2},$ matrix $Q$ $\ $ is the $\pi /4$
rotation matrix. \

In conclusion, the QW on integers with evolution map $\mathcal{E=}$Tr$_{%
\mathrm{c}}$Ad$V_{\mathrm{cl}}$Ad$U,$ randomized by un-bias dichotomous
noise, when treated at the stochastic average level with map $\widehat{%
\mathcal{E}}= $Ad$V_{\mathrm{cl}}\circ \mathcal{E}_{\mathrm{c}},$ becomes
equivalent to a novel non-random QW with map $\widehat{\mathcal{E}}=$Tr$_{%
\mathrm{aux}}$Ad$Y_{\mathrm{cl}}$Ad$Q$. The latter QW, via iterative
relations obeyed by quantum moments of its CP evolution map, exhibits an
eventual transition from a ballistic regime of fast spreading rate to
diffusive regime of slow spreading rate. Similar transition phenomenon
occurs and quantization/randomization interplay is exhibited when
randomization via continuous random variables is introduced.\cite{ebs}

\bigskip \noindent \textbf{Appendix:} \textit{Lemma 1}. \bigskip The
probabilities $p_{i}$ determining the CP map of QW $\mathcal{E}(\rho
_{w})=\sum_{i=\pm }p_{i}E_{i}\rho _{w}E_{i}^{\dagger }$, are given by $%
p_{i}=\left\langle i\right\vert W\circ W^{\ast }\left\vert c\right\rangle $,
where $W$ is an element of the coset $SU(2)/U(1)\thickapprox CP$ coset
corresponding to the reshuffling matrix $U$, and the basis vector $%
\left\vert c\right\rangle $ is determined by the initial state of the coin.

\textit{Proof}. The Hadamard or element-wise product $(A\circ
B)_{ij}=A_{ij}B_{ij}$ of matrices $A,B\in $ $\mathcal{M}(%
\mathbb{C}
^{N}),$ is invariant ($DA\circ (DB)^{\ast }=A\circ B^{\ast }$) under
diagonal $D$ unimodular ($\left\vert D_{ii}\right\vert =1$) matrices. Let $%
V_{q}=V_{\mathrm{cl}}U\otimes \mathbf{1}_{\mathrm{t}}\mathbf{,}$ be a
unitary dilation of CP map $\mathcal{E},$ viz. $\mathcal{E}(\rho _{w})=$Tr$_{%
\mathrm{c}}(V_{q}\rho _{\mathrm{c}}\otimes \rho _{\mathrm{w}}V_{q}^{\dagger
})=\sum_{i=\pm }p_{i}E_{i}\rho _{w}E_{i}^{\dagger },$ where $p_{i}=$Tr$%
(P_{i}U\rho _{c}U^{\dagger })=\left\langle i\right\vert U\circ U^{\ast
}\left\vert c\right\rangle ,$ for $\rho _{\mathrm{c}}=\left\vert
c\right\rangle \left\langle c\right\vert .$ If $U=DW$ be the $SU(2)/U(1)$
coset decomposition of $U,$ for $D\in U(1),$ we obtain $p_{i}=\left\langle
i\right\vert W\circ (W)^{\ast }\left\vert c\right\rangle .$ If \ $D=\exp
(i\phi \sigma _{3}),$ $W\left( \eta \right) =$exp$(\eta \sigma _{+}-\eta
^{\ast }\sigma _{-}),$ we obtain

\begin{equation}
W(\theta )=\left(
\begin{array}{cc}
\sqrt{1-\left\vert z\right\vert ^{2}} & z \\
-z^{\ast } & \sqrt{1-\left\vert z\right\vert ^{2}}%
\end{array}%
\right) =\frac{1}{\sqrt{1+\left\vert \theta \right\vert ^{2}}}\left(
\begin{array}{cc}
1 & \theta \\
-\theta ^{\ast } & 1%
\end{array}%
\right) ,  \label{coset}
\end{equation}%
where $z=\frac{\mathrm{{sin}\eta }}{\eta }=\frac{\theta }{\sqrt{1+\left\vert
\theta \right\vert ^{2}}}\in CP.$ In terms of $p:=(1+\left\vert \theta
\right\vert ^{2})^{-1}=1-\left\vert z\right\vert ^{2}$, the bi-stochastic
matrix $W\circ (W)^{\ast }=p\mathbf{1+}(1-p)\sigma _{1}$ reads%
\begin{equation}
W\circ (W)^{\ast }=\frac{1}{1+\left\vert \theta \right\vert ^{2}}\left(
\begin{array}{cc}
1 & |\theta |^{2} \\
|\theta |^{2} & 1%
\end{array}%
\right) .
\end{equation}

\bigskip

Acknowlegments. The first named author (D. E) is grateful to the Department
of Mathematics, The University of Queensland, for the hospitality extended
to him during a sabbatical stay in which this work was completed.

\bigskip

\end{document}